\documentclass[]{elsarticle}

\usepackage{lineno,hyperref}
\modulolinenumbers[5]

\journal{Physica A}
\usepackage{pdfpages}

\usepackage{amsmath,amssymb}
\usepackage{siunitx}
\usepackage{graphicx}
\usepackage{dcolumn}
\usepackage{bm}
\usepackage{stackrel}
\usepackage{xcolor}

\newcommand{\mm}{Supplementary material}
\newcommand{\mmet}{Supplementary methods}

\newcommand{\la}{\lambda}

\newcommand{\Tw}{\text{Tw}}
\newcommand{\Wri}{\text{Wr}}
\newcommand{\Lk}{\text{Lk}}

\newcommand{\lD}{\ell_D}
\newcommand{\lB}{\ell_B}
\newcommand{\lZ}{\ell_Z}
\newcommand{\lL}{\ell_L}
\newcommand{\lX}{\ell_X}
\newcommand{\ls}{\ell_s}
\newcommand{\lp}{\ell_p}
\newcommand{\hlD}{\hat\ell_D}

\newcommand{\hlX}{\hat\ell_X}

\newcommand{\hlp}{\hat\ell_p}

\newcommand{\Nbp}{N_{\text{bp}}}

\newcommand{\beq}{\begin{equation}}
\newcommand{\eeq}{\end{equation}}
\newcommand{\ben}{\begin{eqnarray}}
\newcommand{\een}{\end{eqnarray}}
\newcommand*{\bfrac}[2]{\genfrac{}{}{0pt}{}{#1}{#2}}

\DeclareRobustCommand{\rchi}{{\mathpalette\irchi\relax}}
\newcommand{\irchi}[2]{\raisebox{\depth}{$#1\chi$}} 

\def\r#1{{\color{black} {#1}}}

\begin{document}

\begin{frontmatter}

\title{A polymer model of bacterial supercoiled DNA including structural transitions of the double helix}

\author{Thibaut Lepage}
\author{Ivan Junier}
\address{CNRS, TIMC-IMAG, F-38000 Grenoble, France}
\address{Univ. Grenoble Alpes, TIMC-IMAG, F-38000 Grenoble, France}

\begin{abstract}
DNA supercoiling, the under or overwinding of DNA, is a key physical mechanism both participating to compaction of bacterial genomes and making genomic sequences adopt various structural forms. DNA supercoiling may lead to the formation of braided superstructures (plectonemes), or it may locally destabilize canonical B-DNA to generate denaturation bubbles, left-handed Z-DNA and other functional alternative forms. Prediction of the relative fraction of these structures has been limited because of a lack of predictive polymer models that can capture the multiscale properties of long DNA molecules. In this work, we address this issue by extending the self-avoiding rod-like chain model of DNA so that every site of the chain is allocated with an additional structural degree of freedom reflecting variations of DNA forms.
Efficient simulations of the model reveal its relevancy to capture multiscale properties of long chains (here up to 21 kb) as reported in magnetic tweezers experiments. Well-controlled approximations further lead to accurate analytical estimations of thermodynamic properties in the high force regime, providing, in combination with experiments, a simple, yet powerful framework to infer physical parameters describing alternative forms.
In this regard, using simulated data, we find that extension curves at forces above 2 pN may lead, alone, to erroneous parameter estimations as a consequence of an underdetermination problem. We thus revisit published data in light of these findings and discuss the relevancy of previously proposed sets of parameters for both denatured and left-handed DNA forms. Altogether, our work paves the way for a scalable quantitative model of bacterial DNA.
\end{abstract}


\end{frontmatter}


\section{Introduction}

In bacteria, and probably in numerous eukaryotes, genome-wide coordination of transcription primarily relies on the physics of DNA in interaction with RNA polymerases~\cite{Junier:2016jj,Junier:2016dj}. Understanding bacterial transcription coordination therefore requires understanding the physics of DNA, more specifically the physics of under and over-wound DNA~\cite{Dorman:1995ha,Hatfield:2002gp,Travers:2005db,Valenti:2011do,Ma:2014jg,Lagomarsino:2015kg}, also known as DNA supercoiling. DNA is indeed continually processed in cells by topoisomerases~\cite{Wang:1991wq}, whose activity allows relaxing the transient constraints generated by replication and by transcription itself, both processes tending to overwind DNA downstream and to underwind it upstream~\cite{Liu1987,Postow:2001vd}. As a result, bacterial DNA is generally underwound, which is commonly referred to as the negative supercoiling of bacterial genomes. In model organisms, and probably in most bacteria~\cite{Junier:2018ar}, this negative supercoiling is further partitioned into approximately 10 kb long topologically independent domains~\cite{Postow:2004fp,Deng:2005bb}.

Although supercoiling is produced at specific sites (e.g.~at the target sites of gyrases), it may impact DNA globally as a consequence of the conservation of the linking number. This linking number, $\Lk$, is equal to the sum of the twist ($\Tw$), the cumulative helicity of the molecule, plus the writhe ($\Wri$), the global intricacy of the molecule~\cite{Strick2003}. Thus, for topologically constrained DNA as in the case of circular molecules (e.g.~plasmids), of topologically constrained linear domains~\cite{Liu1987}, or when DNA molecules are manipulated by magnetic tweezers~\cite{Bryant:2012bi,Lionnet:2012bk,Kriegel:2016hs}, any local variation of the twist results in a global variation of the writhe, and reciprocally. As a consequence, negative DNA supercoiling leads to both superstructuring participating to the global compaction of bacterial chromosomes (through e.g. the formation of so-called plectonemes) and to local modifications of the double helix structure. The latter can induce the formation of functionally important DNA forms different from B-DNA~\cite{Du:2013dda,vol2015book}, including denaturation bubbles~\cite{Benham1979}, cruciforms~\cite{Benham:1982eu}, left-handed Z-DNA~\cite{Rich:1984ec} and left-handed L-DNA~\cite{Bryant2003,Sheinin:2011fv,Oberstrass:2012jc,Vlijm:2015yh}.

One of the most important challenges raised by the biophysical and functional characterizations of bacterial genomes thus consists in predicting the multiscale distribution of supercoiling constraints. Such prediction nevertheless remains challenging, with often no other solution than to resort to numerical simulations~\cite{Irobalevia:2015hj}. In this regard, two main types of DNA polymer models have been proposed. On one hand, explicit polymer models have allowed investigating supercoiling-induced phenomena on the basis of numerical simulations of single DNA chains~\cite{Vologodskii1992,Klenin:1998gq,Vologodskii:2006tn,Ouldridge:2011fz}. On the other hand, phenomenological models~\cite{Marko:2007cf,Meng2014} have led to {\it bona fide} analytical solutions of thermodynamic properties. These have been particularly useful to infer microscopic parameters of alternative DNA forms~\cite{Sheinin:2011fv,Meng2014}, whose knowledge is crucial to both address the exact nature of these forms and parametrize explicit polymer models.

Explicit polymer models can be further divided into two types. Depending on the addressed question, structural details can indeed be coarse-grained at the scale of a single nucleotide~\cite{Ouldridge:2011fz,Manghi2016Ph} or at a larger scale of a few tens base pairs~\cite{Vologodskii1992,Klenin:1998gq,Vologodskii:2006tn,Krajina:2016yg}. Single nucleotide resolution models have thus been useful to investigate properties of small (i.e.~a few hundreds base pairs) DNA molecules as provided by cryo-electron microscopy~\cite{Irobalevia:2015hj,Wang:2017bx} and cyclization data~\cite{vol2015book}, with the possibility to investigate sequence effects in detail (see e.g.~\cite{Mitchell:2017ip} and references therein). Tens base pairs resolution models have instead been useful to address both mechanical and conformational properties of long (i.e.~a few kb or tens kb) B-DNA molecules, taking advantage of much less time-consuming simulations and of the possibility to use parsimonious sets of parameters that can capture bending, torsional and self-avoidance properties of molecules~\cite{Strick2003,vol2015book}. Along this line, the self-avoiding rod-like chain (sRLC) model~\cite{Vologodskii1992} has been paramount to analyze folding properties of positively supercoiled molecules (see below for details of the model). These include molecular extensions~\cite{Vologodskii:1997eb}, torques~\cite{Lipfert2010,Schopflin2012,Lepage:2015gt} and conformation details of superstructures~\cite{Vologodskii1992,Bednar:1994bh,Lepage:2015gt} as measured by magnetic tweezers and cryo-electron microscopy, as well as dynamical properties (see~\cite{Ivenso:2016bp} and references therein) and sequence-dependent phenomena~\cite{Klenin:1995bp,Chirico:1996bi}.

The sRLC model has thus provided a solid framework to infer physical parameters of B-DNA~\cite{Strick2003,Vologodskii:2006tn}. To the best of our knowledge, it has however not yet been adapted to the co-existence of multiple DNA forms, precluding its use to investigate functional properties of bacterial genomes {\it in vivo} and leaving open several important biophysical questions. These include the balance between the global and local relaxations of supercoiling constraints, the exact nature of alternative forms (see e.g.~\cite{Vlijm:2015yh} for a recent discussion about the experimental identification and characterization of denatured DNA (D-DNA) with respect to  L-DNA) and, related to this, the inference of associated mechanical parameters. Namely, while some of the parameters, such as the free energy formation of left-handed DNA forms and of their associated junctions with B-DNA, have been estimated early on in bulk studies~\cite{Benham:1982eu,Rich:1984ec,Singleton:1982tz,Peck:1983gv,Haniford:1983un,Volog:1984cr,Bauer:1993bx}, the estimation of, e.g., the torsional and bending moduli of D-DNA and left-handed forms, has thus far relied exclusively on phenomenological models in the context of single molecule studies~\cite{Marko:2007cf,Sheinin:2011fv,Oberstrass:2012jc,Oberstrass:2013du,Meng2014}. These models, which are based on {\it bona fide} free energy landscape descriptions of the co-existence of multiple DNA forms, have however never been quantitatively assessed, which may explain the existence of strong discrepancies in the prediction of some of the parameters~\cite{Sheinin:2011fv,Meng2014}.

In this work, we aim at filling these gaps by extending the sRLC model so that it includes the possibility to have multiple DNA forms. To this end, we follow an approach initially proposed to tackle the problem of the large flexibility of small DNA molecules~\cite{Cong:2016te}. Namely, as proposed in~\cite{Yan:2004pr,Wiggins:2005tg,Ranjith:2005ph}, the softening of sharply bent DNA can be rationalized by considering a structural degree of freedom allowing the appearance of alternative forms more flexible than B-DNA, with a free energy cost reflecting the destabilization of the double helix. More elaborated models including torsional energies, but still neglecting DNA self-avoidance, have then been proposed to investigate the statistics of DNA denaturation~\cite{Manghi:2009el} and the behavior of stretched DNA molecules~\cite{Efremov:2016pr} under torsional constraints. Models based on torsional energies alone have also been studied in the context of the statistics of DNA denaturation~\cite{Fye:1999vz,Jost:2011hy} as well as to model properties of specific DNA sequences~\cite{Oberstrass:2012jc}. 

Following these studies, here we consider a version of the sRLC model, hereafter referred to as the 2sRLC model, which includes an additional \underline{s}tructural degree of freedom.
Using numerical simulations, we first demonstrate the ability of the 2sRLC model to reproduce magnetic tweezers experiments of long molecules (up to \SI{21}{kb}) for a wide range of stretching forces and supercoiling levels typical of bacterial genomes {\it in vivo}. We next tackle the inference problem of parameters associated with alternative DNA forms. To this end, we circumvent the use of time-consuming simulations by deriving semi-analytical solutions of the thermodynamics when the writhe is negligible and we show that these can be used for any value of supercoiling level and stretching force where plectonemic superstructures are absent. As an application, we discuss the validity of the commonly used phenomenological model proposed by John Marko~\cite{Marko:2007cf} and highlight a previously overlooked underdetermination problem related to the question of parameter inference, which is immanent to the high force analysis of single molecule extension curves. We provide, in this context, our best estimates of parameters associated with D-DNA and L-DNA.

\section{Theory}

\subsection{The discrete sRLC model}

In the sRLC framework, a double-stranded DNA molecule, usually in the form of B-DNA, is modeled as a continuous self-avoiding chain characterized by five fundamental parameters: i) a bending modulus ($K$) defining an associated persistence length ($\lp=K/k_BT$), ii) a torsional modulus ($C$) measured in units of twist persistence length, iii) a winding angle at rest ($\psi$) of the implicitly embedded helix, iv) the distance ($a$) between any two consecutive base pairs, thus defining the number of base pairs per $\lp$, and v) an effective radius ($r_e$) reflecting the hard-core simplification of the electrostatic repulsion of the DNA backbone~\cite{Rybenkov:1997wd}. For B-DNA in physiological conditions ($[\text{NaCl}]\sim\SI{100}{mM}$), $\lp$ typically  lies in $[40-\SI{60}{nm}]$~\cite{Baumann1997,Wenner2002} (in the following, we use $\SI{50}{nm}$), $C$ is on the order of $\SI{100}{nm}$~\cite{Strick1999,Bryant2003}, while $a=\SI{0.34}{nm}$, $\psi=\SI{0.6}{rad/bp}$ and $r_e\approx\SI{2}{nm}$~\cite{Rybenkov:1997wd}.

\begin{figure}
\centering\includegraphics[width=0.9\textwidth]{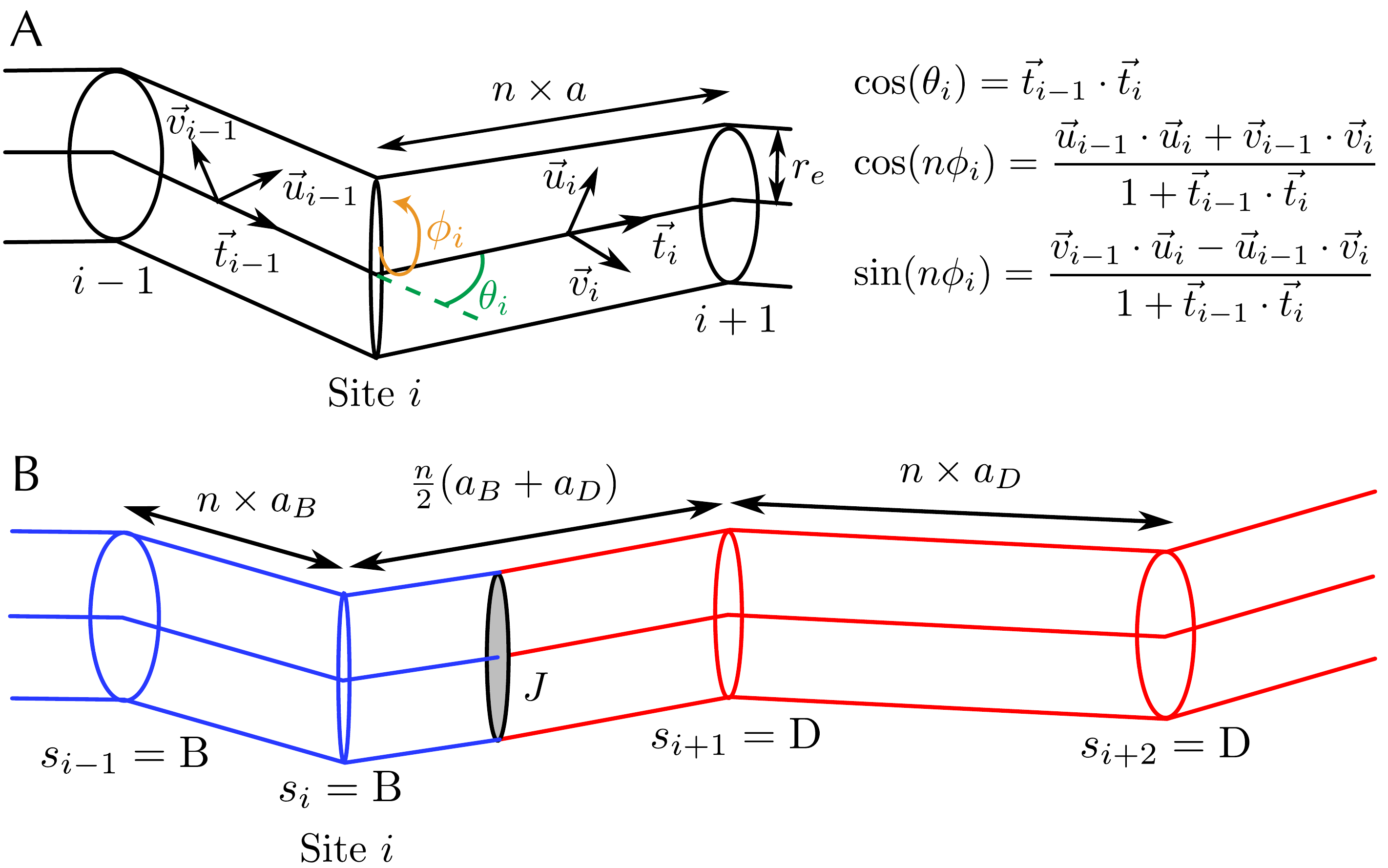}
\caption{{\bf A.} Discrete version of the sRLC: two complete cylinders surrounding a site $i$ are represented along with their local frames $(\vec{t}, \vec{u}, \vec{v})$ used to compute the bending angle $\theta_i$ and the schematically depicted twist angle per bp $\phi_i$. {\bf B.} Extension of the model by allocating a variable $s$ specifying the DNA form of every site. In this case, for symmetry reasons, the length of a cylinder located between sites $k-1$ and $k$ is equal to $n(a_{s_{k-1}}+a_{s_k})/2$, meaning that cylinders may have different lengths because of the different values of $a_s$ characterizing the different forms. Here, we depict the case where sites $i-1$ and $i$ are under the B form (chain in blue), while sites $i+1$ and $i+2$ are denatured (chain in red). We also indicate by a gray disk the domain wall penalty, $J$, associated with the structural cost for transiting from one form to the next.
\label{fig:model}}
\end{figure}

To simulate the folding of a DNA molecule, a discrete version of the model is used, for which the continuous formulation is a good approximation if the discretization is sufficiently fine (see hereafter for further details). Specifically, the chain is divided into identical impenetrable cylinders and an additional parameter, $n$, specifies the number of base pairs per cylinder used in the simulation such that the length of every cylinder is equal to $na$ (Fig.~\ref{fig:model}A) -- in this work, we use $n\leq 10$ such that one cylinder corresponds at most to one B-DNA helix. The energy of any conformation $\mathcal{C}$ of a self-avoiding linear chain in the presence of a stretching force, $f$, then reads:
\beq
E_{\text{sRLC}}(\mathcal{C})=E_B(\mathcal{C})+E_T(\mathcal{C})-fz,
\label{eq:E}
\eeq
where $z$ is the extension of the chain along the axis of the force, while $E_B(\mathcal{C})=\frac{k_BT}{2}\sum_{i=1}^N\frac{\lp}{na}\theta_i^2$ and 
$E_T(\mathcal{C})=\frac{k_BT}{2}\sum_{i=1}^N\frac{nC}{a}(\phi_i-\psi)^2$ respectively stand for the bending and torsional energies. In these equalities, the index $i$ indicates a site around which two consecutive cylinders are articulated, while $\theta_i$ is the bending angle associated with the $n$ base pairs of the site and $\phi_i$ is the winding angle per base pair (Fig.~\ref{fig:model}A); this means that $n\phi_i$ is the torsional counterpart of $\theta_i$ for the site $i$. Note that at the end of the molecule, two additional sites (at $i=0$ and $i=N+1$) set the boundary conditions, which here correspond to a situation where the orientation of the external cylinders is kept fixed along the force axis. Altogether, a linear chain made of $\Nbp$ base pairs thus contains $N+1$ sites and $N$ segments ($N=\Nbp/n=L/na$ where $L$ is the contour length of the molecule). A circular chain (e.g. a plasmid) would contain $N$ sites and $N$ segments, the first site joining the last and first segments together.

Importantly, just as $\theta_i$, $\phi_i$ can be explicitly written as a function of the local frames associated with the surrounding cylinders~\cite{Carrivain2014} (Fig.~\ref{fig:model}A). Using $\Tw\equiv(2\pi)^{-1}\sum_{i=1}^{N}n\phi_{i}$, this ensures that the linking number, $\Lk=\Tw+\Wri$, remains strictly constant for any deformation of a circular chain, provided the chain never crosses itself. $\Lk$ then reflects the topological status of the DNA molecule, for instance a supercoiling constraint if it is different from the corresponding value at rest, $\Lk_0$. For a linear chain, conservation of the linking number further requires the ends of the chain to be attached to two impenetrable walls \cite{Vologodskii:1997eb,LepageJunier}. These respectively play the role of the fixed surface and of the magnetic bead in single-molecule experiments.

\subsection{The discrete 2sRLC model: including multiple DNA forms}

In order to include varying DNA forms at every site $i$ of the sRLC, we follow the lines of denaturation studies~\cite{Fye:1999vz,Yan:2004pr,Wiggins:2005tg,Ranjith:2005ph,Manghi:2009el,Efremov:2016pr} and associate a state variable $s$ with every site ($s=\text{B, D, L or Z}$ if one considers for instance B-DNA, D-DNA, L-DNA and Z-DNA) specifying the associated mechanical parameters: $K_s$ (persistence length $\ls$), $C_s$, $a_s$ and $\psi_s$ -- for simplicity, here we consider a form-independent electrostatic radius $r_e$. In this context, the length of cylinders may vary as a result of both a varying value for $a_s$ and a fixed number of base pairs per cylinder ($n$); here, for symmetry reasons, we consider the length of a cylinder surrounded by two sites with forms $s$ and $s'$ to be equal to $n(a_s+a_{s'})/2$ (Fig.~\ref{fig:model}B).

In addition to form-dependent mechanical parameters, alternative DNA forms come along with free energy formation costs on the order of one $k_BT$ per bp~\cite{SantaLucia:1998uz,Huguet:2010gs}, reflecting the internal deformation of base pairing and base pair stacking. Here, we denote this cost by $\gamma_s^0$ and set $\gamma_B^0=0$ (reference form). Finally, a domain wall penalty, $J$, must be considered between any two sites having different states. This penalty accounts for the internal free energy cost to transit from one DNA form to another one~\cite{Peck:1983ea,Manghi:2009el,Oberstrass:2013du} (Fig.~\ref{fig:model}B) -- in effect, it constrains alternative forms to gather in as few domains as possible. This wall penalty is in principle associated with two consecutive base pairs and therefore includes an entropic contribution on the order of $\ln(n)$ coming from the $n$ possible choices of these base pairs. For simplicity, here we consider this contribution as already included in the values of $J$.

Altogether, the energy of a self-avoiding 2sRLC conformation $\mathcal{C}$ having an arbitrary sequence $\{s_i\}_{i=1..N}$ of DNA forms reads:
\beq
E_{2sRLC}(\mathcal{C})=E_\gamma(\mathcal{C})+E_B(\mathcal{C})+E_T(\mathcal{C})+E_J(\mathcal{C})-fz,
\label{eq:Es}
\eeq
where $E_\gamma(\mathcal{C})=n\sum_{i=1}^N\gamma_{s_i}^0$ is the total free energy formation costs coming from alternative forms,
$E_B(\mathcal{C})=\frac{k_BT}{2}\sum_{i=1}^N\frac{\ell_{s_i}}{na_{s_i}}\theta_i^2$ is the bending energy, 
$E_T(\mathcal{C})=\frac{k_BT}{2}\sum_{i=1}^N\frac{nC_{s_i}}{a_{s_i}}(\phi_i-\psi_{s_i})^2$ is the torsional energy and $E_J(\mathcal{C})=J\sum_{i=1}^{N-1}(1-\delta_{s_i,s_{i+1}})$
is the total wall penalty, with $\delta_{s_i,s_{i+1}}=1$ if $s_i=s_{i+1}$, $0$ otherwise.

\section{Results and discussion}

\subsection{Effective 2sRLC model}

Solving the thermodynamics of the 2sRLC model is challenging because of the self-avoidance constraints. Yet, following upon previous studies on single DNA forms~\cite{Gebe:1995eu} and on the statistics of DNA denaturation~\cite{Fye:1999vz,Jost:2011hy}, it is possible to integrate out the torsional degrees of freedom (the $\phi_i$'s) under the constraint of a fixed $\Lk$. This leads to an equivalent effective model that has the benefit to offer much better simulation performances~\cite{Vologodskii:1997eb,Lepage:2015gt} and further analytic treatment (see below). In the case of the 2sRLC model, the resulting conformational energy can be decomposed as in Eq.~\ref{eq:Es}, but with an effective free energy of formation and an effective torsional energy that respectively read (see \mmet~in\mm~for details):
\begin{eqnarray}
E_\gamma^{eff}(\mathcal{C})&=& n\sum_{i=1}^N\gamma_{s_i}\text{ where $\gamma_{s_i}=\gamma_{s_i}^0-\frac{1}{2n}\ln\frac{a_{s_i}}{C_{s_i}}$}
\label{eq:Egeff}
\\
E_T^{eff}(\mathcal{C})&=&
\frac{k_BT}{2} \left[\frac{\Nbp}{\rchi(\mathcal{C})}\left(2\pi\frac{\Tw(\mathcal{C})}{\Nbp}-\psi(\mathcal{C})\right)^2+\ln(\rchi\left(\mathcal{C})\right)\right]
\label{eq:ETeff}
\end{eqnarray}
where, again, $\Nbp=nN$ is the length of the molecule in base pairs, $\psi(\mathcal{C}) \equiv \sum_{i=1}^{N}\psi_{s_i}\big/N$ and $\rchi(\mathcal{C}) \equiv \sum_{i=1}^{N}\frac{a_{s_i}}{C_{s_i}}\big/N$ are {\it conformation-dependent} average quantities, namely, the average twist angle per bp at rest and the average torsional susceptibility per bp.

Eq.~\ref{eq:ETeff} thus generalizes to the case of any number of different DNA forms the possibility, for topologically constrained DNA molecules, to obtain an equivalent effective model with a torsional energy quadratic in $\Tw$. Note, nevertheless, that here the associated parameters depend on the composition of the molecule in these DNA forms. In the particular case of a single (B-DNA) form, one for instance recovers $\psi(\mathcal{C})=\psi_B$ and $\rchi(\mathcal{C})=a_B/C_B$ for all conformations, such that $E_{T}^{eff}(\mathcal{C})=\frac{k_BTC_B\Nbp}{2a_B}\left(2\pi\frac{\Tw(\mathcal{C})}{\Nbp}-\psi_B\right)^2$ as demonstrated in~\cite{Gebe:1995eu}.

\subsection{Capturing the phenomenology of negatively supercoiled DNA}

To demonstrate the ability of the 2sRLC model to capture the phenomenology of negatively supercoiled DNA, we performed Monte-Carlo simulations of long chains using the above effective energies. To this end, we have adapted the standard Monte-Carlo procedure used to simulate the equilibrium folding of the sRLC model~\cite{Vologodskii:2006tn} in order to include the possibility for mechanical parameters to change at any site of the chain, at any iteration of the simulation (see~\cite{LepageJunier} for details). In practice, it involves a new transition type so that for any site $i$ independently of the other sites, the value of the variable $s_i$ may change under the constraint of detailed balance. 
\r{Corresponding variations of cylinder lengths (Fig.~\ref{fig:model}B) are thus managed as other elementary Monte-Carlo moves. In particular, if they lead to a collision, the transition is rejected. More generally, length variation of a cylinder always requires an adjustment of the positions and orientations of the surrounding cylinders to preserve the continuity of the chain. As explained in~\cite{LepageJunier}, this is realised by performing specific random rotations that are constrained by the geometry of the chain.}

\begin{figure}
\centering\includegraphics[width=\textwidth]{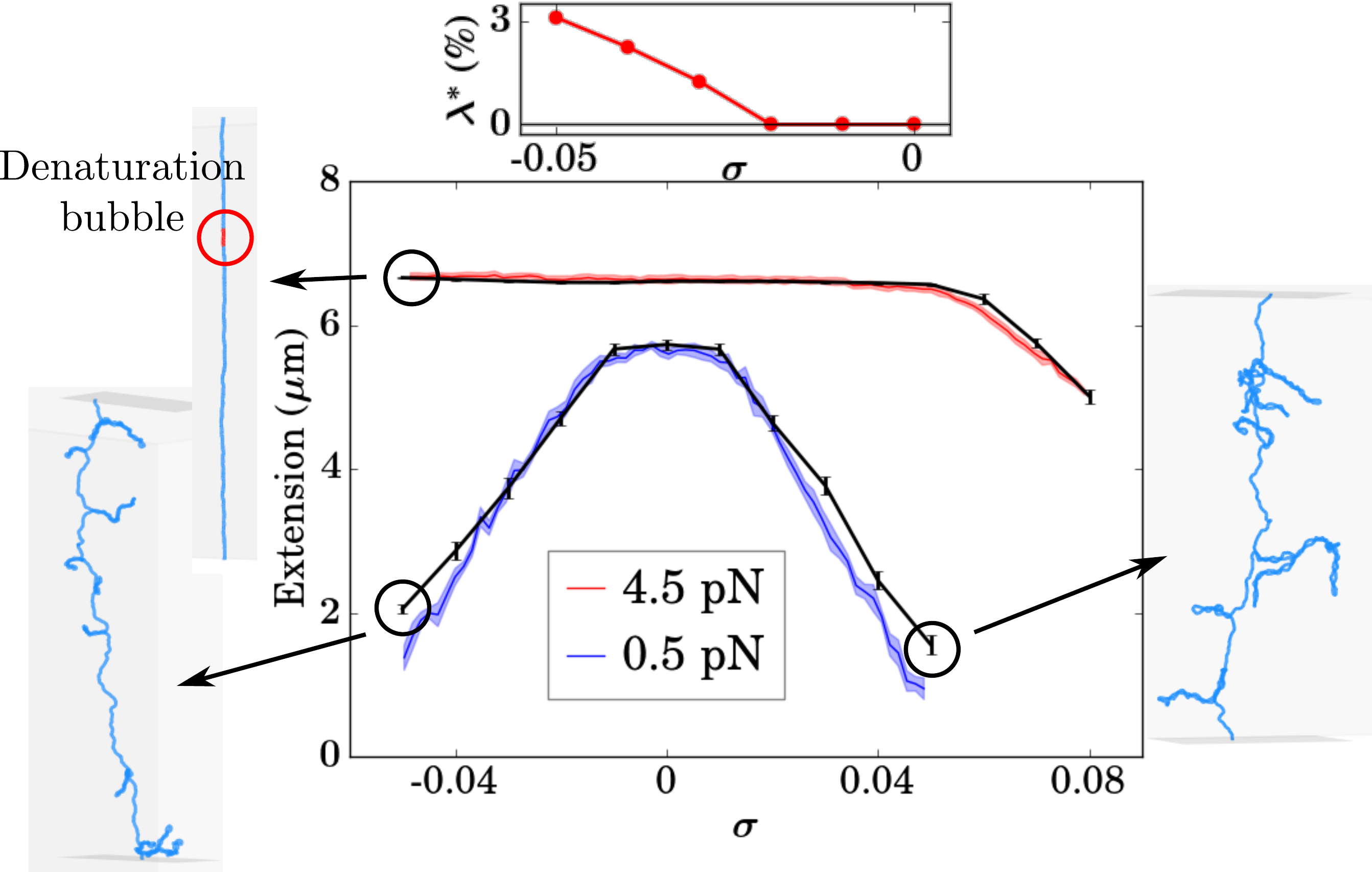}
\caption{Extension of a 21~kb molecule as a function of $\sigma$, for different forces, obtained experimentally (in red and blue, data from~\cite{Vlijm:2015yh}) and numerically using a 2sRLC model that includes the possibility to have B-DNA plus an alternative DNA form along the chain (see main text for values of the microscopic parameters). Error bars and colored areas represent one standard deviation due to thermal fluctuations. At low forces, the curves are symmetrical with respect to $\sigma$, with the formation of branched plectonemic structures at large values of $\vert\sigma\vert$ as indicated by the typical conformations observed in the simulations (leftmost and rightmost panels).  At high forces, the buckling transition toward the plectonemic regime occurs only for positive supercoiling. For negative supercoiling, denaturation bubbles appear instead (upper left panel, in red). Upper panel: corresponding estimation of the equilibrium fraction $\lambda^*$ of the alternative form as a function of $\sigma$.}
\label{fig:simus}
\end{figure}

Using this Monte-Carlo method, we could quantitatively reproduce extensions of several kilo base pairs long molecules as obtained in magnetic tweezers experiments for a wide range of forces and supercoiling densities $\sigma=(\Lk-\Lk_0)/\Lk_0$. As an example, we report in Fig.~\ref{fig:simus} both our simulation results and the experimental results of Cees Dekker's lab~\cite{Vlijm:2015yh} for a 21~kb molecule, for $\sigma \in [-0.05,0.05]$, $f=\SI{0.5}{pN}$ and $f=\SI{4.5}{pN}$, providing as a bonus a prediction of the equilibrium fraction of alternative forms present along the molecule (top panel). DNA denaturation being expected to be the most likely alternative form at these forces and supercoiling densities~\cite{Oberstrass:2012jc,Vlijm:2015yh}, here we considered the situation where only B-DNA and D-DNA could form, with the following parameters leading to good agreement with experimental curves: $a_D=\SI{0.45}{nm}$, $\lD=\SI{15}{nm}$, $C_D=\SI{10}{nm}$ and $\gamma_D=2k_BT$ (see below for discussion of these values) -- we intentionally used a $\lD$ much larger than the $\SI{4}{nm}$ diameter of the chain because smaller values would lead, in any case, to larger effective persistence lengths~\cite{Barrat:1993el}. We used $J=10k_BT$, in accord with previous analyses~\cite{Peck:1983ea,Oberstrass:2012jc,Oberstrass:2013du}, and considered a discretization level of $n=10$, leading to $\simeq 3.3$ cylinders per $\lD$ and 15 cylinders per $\lB (=\SI{50}{nm})$, a value that both offers reasonable computational times and ensures properties of B-DNA superstructuring to be insensitive to discretization procedures~\cite{Lepage:2015gt,LepageJunier}. Finally, the quantitative reproduction of the buckling transition at $\sigma >0$ (right part of the curves in Fig.~\ref{fig:simus}) required us to use $C=\SI{80}{nm}$ at $f=\SI{4.5}{pN}$ and $C=\SI{100}{nm}$ at $f=\SI{0.5}{pN}$ (Fig.~S1). This is in accord with the observed systematic discrepancies occurring at low forces between sRLC and experimental extensions~\cite{Nomidis:2017ii}, a consequence of a coupling between torsional and bending deformations stemming from the asymmetry of the DNA helix~\cite{MarkoSiggia}. 

Our simulations corroborate well-known results of magnetic tweezers experiments, with notably the presence of two regimes (Fig.~\ref{fig:simus}): i) a low force regime ($f\lesssim\SI{0.5}{pN}$) that is symmetric under the transformation $\sigma\to-\sigma$ and where torsional stress is partially relaxed through the formation of plectonemes, some of them being branched, and ii) a $\sigma$-asymmetric high force regime ($f\gtrsim\SI{2}{pN}$) where negative torsional stress is partially relaxed by forming denaturation bubbles. For intermediate forces at $\sigma<0$, both conformations with denaturation bubbles and plectonemic conformations become locally stable, being separated by high free energy barriers (Fig.~S2). This is in accord with the experimental findings, in this regime, of large equilibrium fluctuations of the extension ~\cite{Salerno2012,Meng2014}, which have been shown, in the case of a 2.4~kb long molecule, to reflect a two-state behavior between plectonemic conformations and conformations with D-DNA bubbles~\cite{Meng2014}. In our simulations of comparable 2.4~kb long molecules, this manifests through an hysteresis cycle in the force-extension diagram (Fig.~S2), with the drawback that specific numerical methods, such as metadynamics~\cite{Laio:2008ij}, must be developed in this regime to get access to thermodynamic properties, \r{including the height of the free energy barriers}.

Our simulations also corroborate the existence of conformations where denaturation bubbles locate at the apex of plectonemes (Fig.~S3). These have been predicted to occur using a polymer model of DNA coarse-grained at the nucleotide level~\cite{Matek:2015gb}. Here, we observe that these tip-bubble conformations, as coined in~\cite{Matek:2015gb}, are all the more favored that the system is close to the "spinodal" point associated with the plectonemic state -- the force at which the free energy barrier separating pure plectonemic states from states with denaturation bubbles becomes on the order of $k_B T$.

Most importantly, while trying to reproduce previously published experimental extensions, we have found that different sets of parameters could lead to similar results. For instance, the D-DNA persistence length used in Fig.~\ref{fig:simus} is typically 5 times larger than that previously estimated ~\cite{Marko:2007cf,Sheinin:2011fv}, while previous works have led to values of the torsional modulus that may differ by more than 20-fold~\cite{Sheinin:2011fv,Meng2014}.
Overall, this raises the question of the extent to which it is possible to infer mechanical parameters of alternative forms using extension curves alone as obtained in magnetic tweezers experiments, what we discuss now in detail.

\subsubsection{Zero writhe approximation in the high force regime}

One could in principle use the above numerical simulations to search for sets of parameters that best reproduce extension curves.
However, independently of the efficiency of simulations, an exhaustive exploration of the 5 parameters ($\lD$, $C_D$, $a_D$, $\psi_D$ and $\gamma_D^0$) and the domain wall penalty $J$ is extremely challenging. Moreover, the high free energy barriers separating plectonemic conformations from conformations with D-DNA bubbles precludes obtaining the relative fraction of each state at equilibrium using standard Monte-Carlo methods. Curves at intermediate forces between $\SI{0.5}{pN}$ and $\SI{2}{pN}$ cannot thus be currently exploited to further constrain parameters as previously done in the context of phenomenological models~\cite{Meng2014}.

To partially circumvent these problems, we focus on the regime of high forces, for which the thermodynamics can be solved in the approximation of both a negligible writhe, i.e. when $\Lk=\Tw$, and negligible self-avoidance properties (see below for the range of forces where this approximation holds). In this approximation, it is indeed possible to obtain an explicit formula for the free energy profile of the system, $F_{f,\sigma}(\{\la_i\})$, as a function of the fraction of each of the possible alternative forms at a given stretching force $f$ and supercoiling density $\sigma$. In the following, for simplicity we discuss the case of two forms, with in mind B-DNA plus a certain fraction $\lambda$ of an alternative form, hereafter denoted as X-DNA -- the generalization to more than two forms is straightforward (see \mmet). 

Calling $\mathcal{J}(\la)$ the free energy associated with the multiple possibilities to distribute the $\la N$ sites of the alternative form into distinct domains (see SI for its derivation), $g_{s,f}(L_s)$ the free energy of the WLC at force $f$ associated with the form $s$ with contour length $L_s$, and using the notations $\rchi_s=a_s/C_s$ and $\rchi/\psi(\la)=(1-\la)\rchi_B/\psi_B+\la\rchi_s/\psi_s$, we obtain (\mmet):
\begin{eqnarray}
\nonumber
F_{f,\sigma}(\la)&=&
\underbrace{\la \Nbp\gamma_X+\mathcal{J}(\la)}_{\bfrac{\text{free energy formation}}{\text{of alternative domains}}}
+\underbrace{g_{X,f}(\la \Nbp a_X) +g_{B,f}((1-\la)\Nbp a_B)}_{\text{bending}+\text{stretching}}\\ &&+\underbrace{\frac{k_BT}{2}\left[\frac{\Nbp}{\rchi(\la)}\left(1+\sigma-\frac{\psi(\la)}{\psi_B}\right)^2+\ln\left(\rchi(\la)\right)\right]}_{\text{torsion}}
\label{eq:F}
\end{eqnarray}

Using this free energy profile, equilibrium properties of the 2sRLC model in the zero writhe approximation can be computed by performing numerical integration  over $\la$ of Boltzmann-weighted observables \r{(see Eq.~10 in \mmet~and subsequent explanations)} with, here, a particular interest in equilibrium values of the extension, $z^*=-\langle \partial_f F_{f,\sigma}(\la)\rangle_\la$, the torque $\Gamma^*=
(2\pi\Tw_0)^{-1}\langle \partial_\sigma F_{f,\sigma}(\la)\rangle_\la$ (see \mmet~for derivation and further details), and the fraction of alternative form, $\la^*=\langle \la\rangle_\la$, where $\langle \bullet\rangle_\la\equiv\int d\la\bullet \exp\left[-F_{f,\sigma}(\la)/k_BT\right]\big/\int d\la\exp\left[-F_{f,\sigma}(\la)/k_BT\right]$. We also compute the equilibrium number of domains, $X^*$, using a free energy surface in the $(\la,X)$ space (Eq.~6 in \mmet).

\subsubsection{Validity domain of the zero writhe approximation}

To assess the validity domain of the zero-writhe approximation, we compare the above Boltzmann-weighted equilibrium values to those obtained in Monte-Carlo simulations of the corresponding 2sRLC model. As a case study, we investigate a 1~kb long molecule with an alternative form characterized by $\lX=\SI{15}{nm}$, $a_X=\SI{0.54}{nm}$, $\psi_X=0$, $C_X=\SI{10}{nm}$ and $\gamma_X=1.6kT$, having in mind a form close to D-DNA; just as in the analysis of Fig.~\ref{fig:simus}, the relatively high value of $\lX$ compared to previous estimations for D-DNA  ($3-\SI{4}{nm}$~\cite{Marko:2007cf,Sheinin:2011fv}) was chosen for practical purposes, that is, to avoid potential artifacts that may occur in the simulations when the persistence length is smaller or on the order of $r_e$~\cite{LepageJunier} and to avoid considering (longer) effective persistence lengths in our analytical approximations~\cite{Barrat:1993el}. We also used a fine discretization level of $n=4$ to get a large number of cylinders per $\lD$ ($\simeq 7$) so that to prevent any discrepancy coming from the discretization procedure of the rod-like chain.

\begin{figure}
\centering\includegraphics[width=0.9\textwidth]{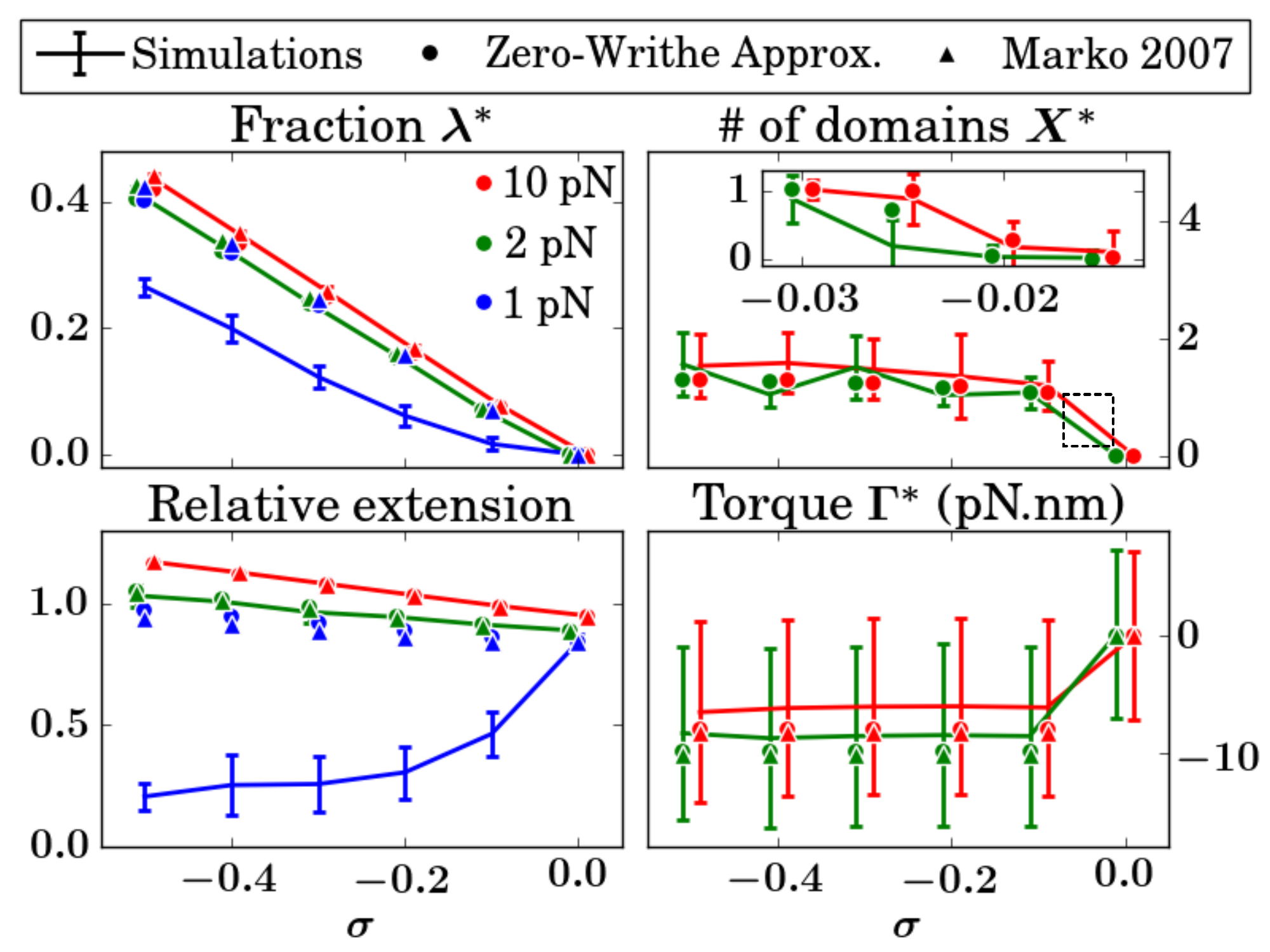}
\caption{Comparison between numerical simulations, the zero writhe approximation and Marko's phenomenological approach for the equilibrium value of different quantities of interest as a function of the negative supercoiling: the fraction $\lambda^*$ of denatured alternative form, the number $X^*$ of alternative domains, the torque $\Gamma^*$ and the extension of the molecule (relative to its contour length under the single B form).
Error bars represent one standard deviation due to thermal fluctuations; in particular, this reveals that torque fluctuations are large.
For readability, the green points (2~pN) and red points (10~pN) have respectively been slightly shifted left and right with respect to the exact values of $\sigma$: $0, -0.1, -0.2, -0.3, -0.4, -0.5$ ($-0.015,-0.02,-0.025,-0.03$ in the inset of $X^*$).
The transition between fully stretched conformations and plectonemic conformations (where the approximation does not hold anymore) occurs between 1 and \SI{2}{pN}. \r{For all panels: the apparent absence of error bars implies that they are smaller than symbols.}}
\label{fig:MvsNoW}
\end{figure}

Regarding $\lambda^*$ and the extension, we observe an excellent agreement between simulation results and the zero-writhe approximation for any force above $\SI{2}{pN}$ (left panels in Fig.~\ref{fig:MvsNoW}), with a relative error lower than 5\% (Fig.~S4).
For the number of domains the error is on the order of 20\% (a higher error is expected as a consequence of the discrete nature of $X$), and up to 30\% for the torque (side effect of the error in $X^*$).
Most remarkably, this agreement is observed for a wide range of supercoiling densities, from the transition point at physiological $\sigma\simeq -0.025$ associated with the onset of the first denaturation bubble (see inset in upper right panel of Fig.~\ref{fig:MvsNoW}) to large non-physiological negative values of $\sigma=-0.5$, and for forces down to the onset of plectonemic superstructures. The strong deviation observed at $\SI{1}{pN}$ is indeed the result of a buckling transition that occurs in the simulations, which is not captured by the zero writhe approximation. In other words, the zero writhe approximation provides a good description of the system whenever plectonemic superstructures are absent, paving the way for an efficient method for the quantitative inference of physical properties of alternative DNA forms in the high force regime (see below). In this regard, we note that the inclusion of finite size corrections~\cite{Seol2007} for computing the WLC terms $g_{X,f}$ and $g_{B,f}$ leads to very similar results (Fig.~S5), suggesting the possibility to use the classical infinite length approximation of the WLC~\cite{Bouchiat1999}, although the fraction of alternative form may be large.

We finally used the same simulations to assess the validity of the commonly used phenomenological approach proposed by John Marko a decade ago~\cite{Marko:2007cf}. In a nutshell, Marko proposed a model with two separated phases (B-DNA and X-DNA), each of them being allocated with a certain supercoiling density and a free energy that is quadratic in the latter; the two supercoiling densities are coupled because of the conservation of the linking number. Then, supposing a linear relationship between the fraction $\lambda$ of X-DNA and the total supercoiling density ($\sigma$) and considering the torques to be equal in both phases, Marko used a free energy minimization procedure to compute the supercoiling densities specific to each phase, leading back to the computation of the free energy as a function of $\sigma$. Here, our analysis shows that this phenomenological approach provides as well a remarkable description of the 2sRLC equilibrium properties (Fig.~\ref{fig:MvsNoW}), excluding the number of domains, $X$, which is not a variable of the model~\cite{Marko:2007cf}.

\subsection{Inferring mechanical parameters of alternative forms from extension curves}

Alternative DNA forms mostly occur in presence of B-DNA. Estimation of their parameters, which ultimately reflect statistical properties of the dynamics of strands, has thus been a matter of debate. Both bulk~\cite{SantaLucia:1998uz} and single-molecule~\cite{Huguet:2010gs} measurements have nevertheless converged to free energy formations of D-DNA on the order of one $k_BT$ per bp. Cyclization measurements of small circular DNA molecules have also revealed an independence of D-DNA strands, suggesting to use $\phi_D=0$~\cite{Kahn:1994tf}. A zero-writhe phenomenological modeling of these experiments have then led to $a_D=\SI{0.54}{nm}$, $\lD \simeq 2\text{ to }3~\text{nm}$ and $C_D=\SI{1}{nm}$~\cite{Kahn:1994tf}. In contrast, investigation of the co-existence of D-DNA with B-DNA at forces $\lesssim \SI{1}{pN}$ and $|\sigma| \lesssim 0.06$, using a phenomenological model of the coexistence of twisted B-DNA, plectonemic B-DNA, and D-DNA, has led to $C_D=\SI{28}{nm}$~\cite{Meng2014}. Regarding left-handed DNA forms, static and dynamic laser light scattering experiments of Z-DNA solutions led to $\lZ\simeq\SI{200}{nm}$~\cite{Thomas:1983ur}, 
whereas a best fitting procedure using Marko's phenomenological model revealed a flexible L-DNA form, with $a_L=\SI{0.48}{nm}$ and $\lL=\SI{3}{nm}$~\cite{Sheinin:2011fv}, and a torsional modulus $C_L$ lying between $10$ and $\SI{20}{nm}$. Similar values of $C_L$ were found by investigating the collapse of “pure” L-DNA molecules ~\cite{Oberstrass:2012jc}. Finally, location of the buckling transition of such "pure" L-DNA molecules at high forces led to $\psi_L=\SI{-0.5}{rad/bp}$~\cite{Oberstrass:2012jc,Vlijm:2015yh}.

Here, we aim at using the predictive power of our zero writhe approximation of the 2sRLC model (Fig.~\ref{fig:MvsNoW}) in order to infer some of the parameters of alternative forms. Specifically, we discuss the possibility, as proposed in~\cite{Sheinin:2011fv}, to estimate parameters of D-DNA and L-DNA using $\sigma$-extension curves, alone, in the high force regime. To this end, we first test our capacity to recover original parameters using simulated data and, then, discuss estimations from real data. 

\subsubsection{Simulated data: assessing parameter inference in an underdetermination context}

To test our capacity to recover original parameters (hereafter indicated by a hat, e.g., $\hat a_X$) from simulated $\sigma$-extension data, we use as a generative model the 2sRLC model under the zero writhe approximation (Eq.~\ref{eq:F}) and question the possibility to recover original parameters using the same generative model. Regarding the inference method, we follow the procedure of~\cite{Sheinin:2011fv}, which consists in determining optimal parameters such that extensions of the model best fit experimental extensions over a wide range of forces and supercoiling densities (\mmet).

\begin{figure}
\centering\includegraphics[width=\textwidth]{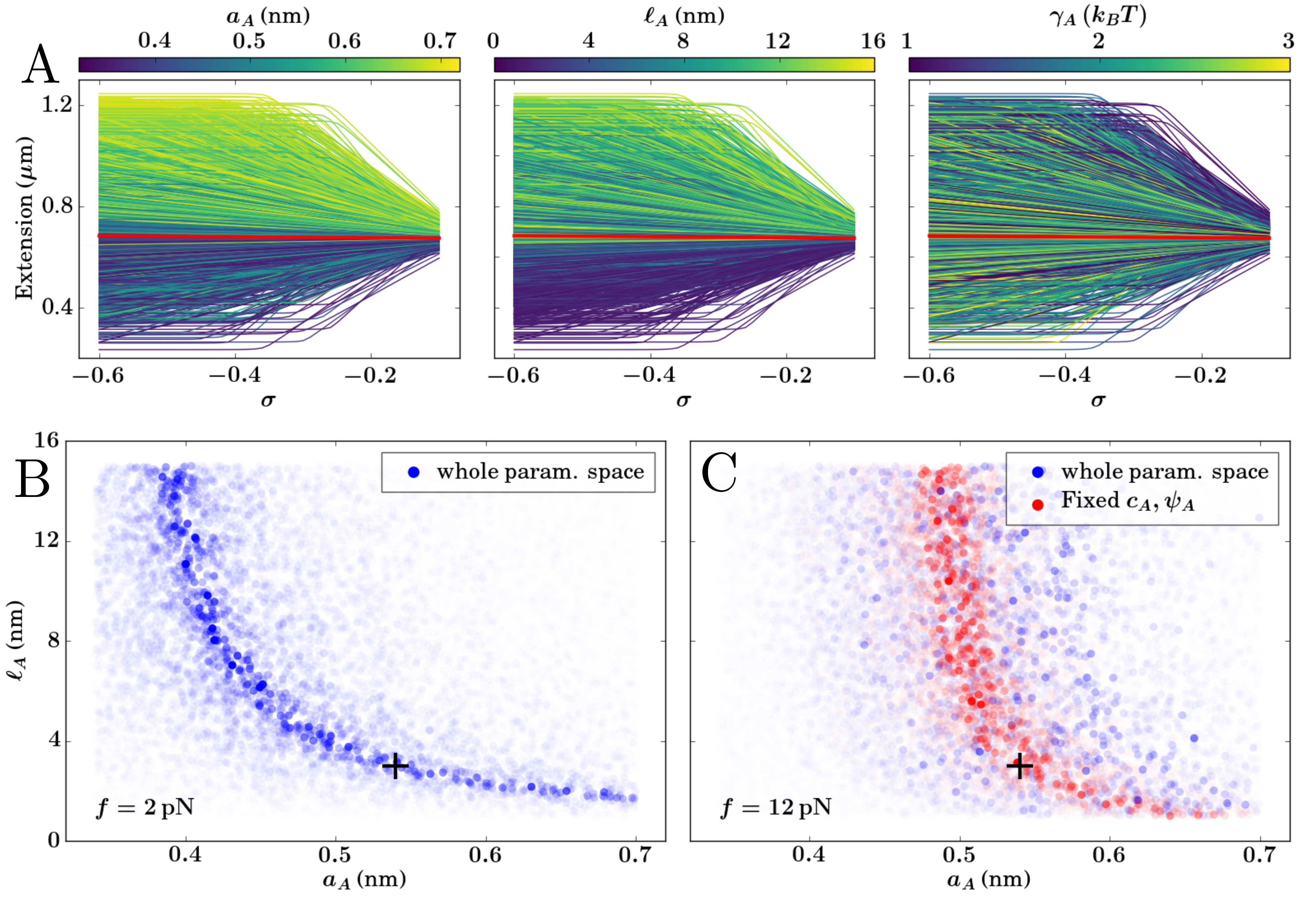}
\caption{Test of the possibility to recover parameters of an alternative X-DNA co-existing with B-DNA, using simulated extension curves (red curves in panel A). To this end, we first generated $\sigma$-extension curves using the zero writhe approximation model of Eq.~\ref{eq:F} with a specifically predefined X-DNA (see main text for parameters). We next performed an inference procedure using the same generative model by producing $10\,000$ different curves corresponding to a random combination of the 6 parameters of X-DNA.
\textbf{A}: $\sigma$-extension curves at $f=\SI{2}{pN}$ where the color of each curve is given by the value of one of the parameters: $a_X$ (left), $\lX$ (center) and $\gamma_X$ (right). See Fig.~S6 for the other parameters. We can see that, compared to $\gamma_X$, $a_X$ and $\ell_X$ have a well-defined, strong influence on extension curves. In these cases, the curves indeed appear to be sorted as a function of the values of the parameters.
\textbf{B}: Each point corresponds to a random set of parameter values, with $x$- and $y$-axes corresponding to the values of $a_X$ and $\lX$, respectively.
The opacity is inversely correlated (using an arbitrarily scale) to the root-mean-square deviation (RMSD) of the curves in A with respect to the red one, so that only the best fits are visible.
We observe a crest of small RMSD values that go through the original parameters $(\hat a_X, \hlX)$ (black cross).
\textbf{C}: At a higher force (\SI{12}{pN}), the front is much less visible and the original set of solutions becomes isolated among a broad cloud of RMSD values (in blue). The crest becomes clearer again by setting $\psi_X=\hat\psi_X(=0)$ and $C_X=\hat C_X(=\SI{10}{nm})$ (in red). The intersection between crests obtained at different forces then allows for an accurate estimation of $\hat a_X$ and $\hlX$ (Fig.~S7).
}
\label{fig:testinf}
\end{figure}

First, compared to the other parameters, we find that $a_X$ and $\lX$ (or equivalently $K_X$) have a more systematic impact on $\sigma$-extension curves. More precisely, using a large random set of parameters, the various $\sigma$-extension curves appear to be sorted according to the values of both $a_X$ and $\lX$, whereas they appear to be randomized with respect to the values of the other parameters (Fig.~\ref{fig:testinf}A, Fig.~S6) -- we nevertheless note that best fits correspond to values of $\psi_X$ that are distant from $\psi_B$ (Fig.~S6). The goodness of fit can be further quantified by computing the root-mean-square deviation (RMSD) of each tested curve with respect to the original curve. Reporting these RMSDs in the plane $(a_X,\lX)$ at a given force then reveals the existence of a well-defined continuous set of parameters with similar locally optimal solutions and going through $(\hat a_X,\hlX)$, hereafter called a crest of solutions (Fig.~\ref{fig:testinf}B) -- note that these solutions are always suboptimal with respect to the original set of parameters. The overall shape of this crest reflects the existence of two types of solutions: 1) solutions  corresponding to almost straight pieces of X-DNA (large $\lX$, the exact value being irrelevant); 2) solutions corresponding to more disordered domains (small $\lX$) with longer double helices (larger $a_X$). Most importantly, an analysis of solutions in the vicinity of $(\hat a_X,\hlX)$ corroborates the existence of a large spectrum of possible values for $J$, $\gamma_X$, $\psi_X$ and $C_X$ (Fig.~S8), meaning that these parameters cannot be estimated using a single extension curve, i.e., a single force. Moreover, the presence of the crest implies the existence of some degeneracy that precludes the unambiguous determination of $\hat a_X$ and $\hlX$, too. 

In principle, the value of $\hat C_X$ could be addressed using torque measurements~\cite{Oberstrass:2012jc}, yet only if one has access to $\la(\sigma)$ (see \mmet). For $\hat a_X$ and $\hlX$, one might expect that the underdetermination problem akin to a single force could be resolved by considering multiple forces. However, we find that, without fixing any other parameters, the crest gets blurrier as the force increases (blue points in Fig.~\ref{fig:testinf}C). To understand this intriguing result, we performed the same analyses but by setting one arbitrary parameter to its right value. We found that fixing either $J$ or $\gamma_X$ leads to the same underdetermination problem. In contrast, fixing either $\psi_X$ or $C_X$ makes a crest of locally suboptimal solutions re-appear. However, setting $\psi_X$ to the right value $\hat\psi_X=0$ results in a crest that is off the right solution $(\hat a_X,\hlX)$ (Fig.~S9), a shift that disappears only by setting $C_X=\hat C_X$ (red points in Fig.~\ref{fig:testinf}C).

Overall, these results show that inference of $\hat a_X$ and $\hlX$ from extension curves requires to have some prior knowledge about the nature of the alternative form, more particularly about its torsional parameters ($\hat \psi_X$ and $\hat C_X$). This may be explained by the fact that the (effective) torsional energy is quadratic in $\sigma$, such that any parameter that affects this energy is expected to have a large impact on any curve plotted as a function of $\sigma$. When the values of $\hat\psi_X$ and $\hat C_X$ are known, a minimization over all forces then lead to a fair estimation of both $\hat a_X$ and $\hlX$ which is all the more accurate that the range of available forces is large (the crossing of the different crests of solutions becoming well-defined as shown in Fig.~S7).

\subsubsection{Experimental data: estimation of D-DNA and L-DNA parameters}

\begin{figure}
\centering\includegraphics[width=0.7\textwidth]{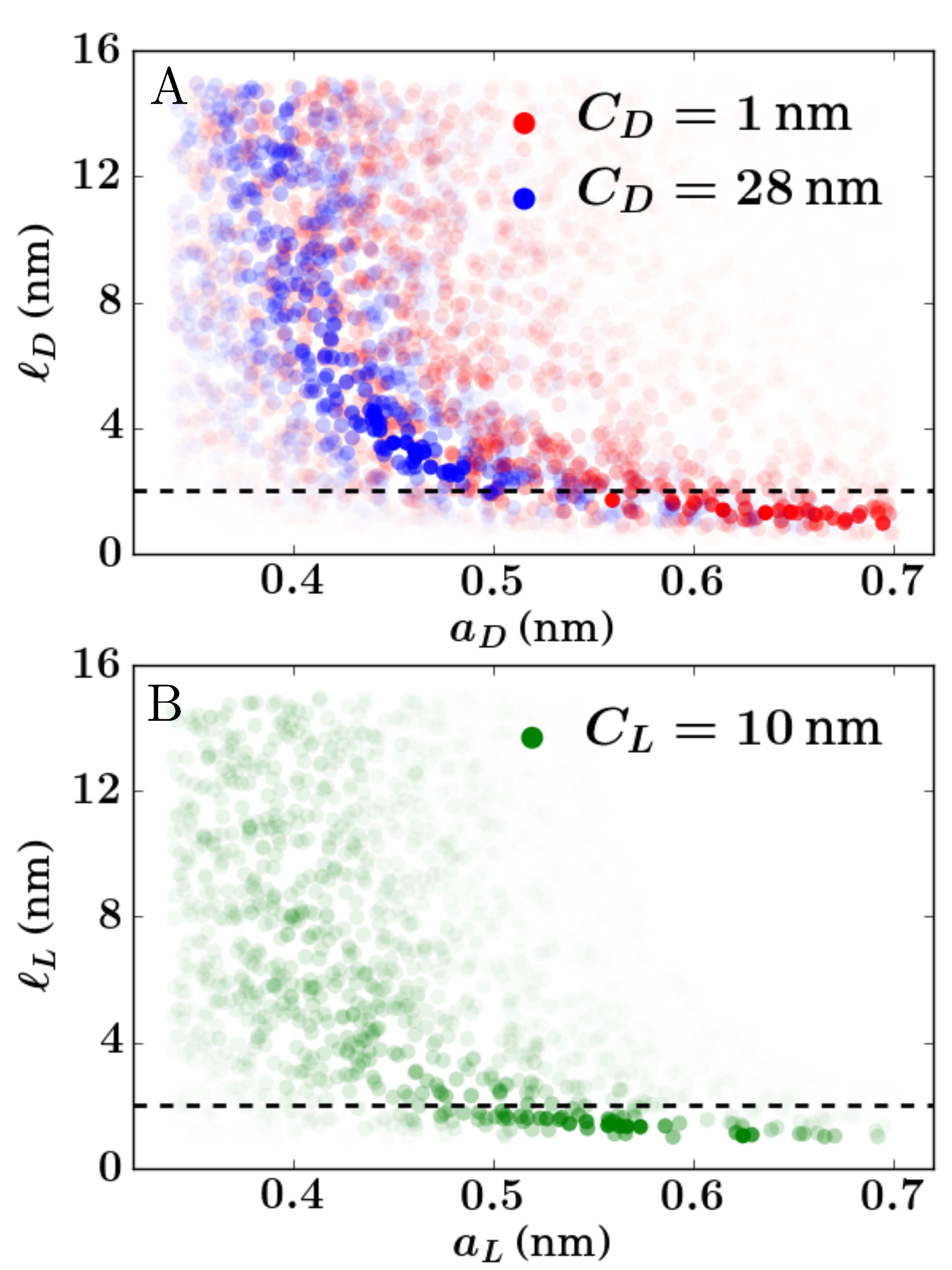}
\caption{\textbf{A}: Inference of D-DNA parameters using experimental data from~\cite{Sheinin:2011fv}. Each point represents the cumulative (over several forces ) RMSD resulting from a best fit procedure of experimental extension curves using numerical curves of the 2sRLC model generated under the zero-writhe approximation (see Eq.~\ref{eq:F}) and for a given set of parameters. Points are all the darker that the cumulative RMSD is small (arbitrary scale). For each set of parameters, the cumulative RMSD was computed for $\sigma$ varying in $[-0.6,-0.1]$ and over $6$ different forces ranging from $2.5$ to $\SI{12}{pN}$. To mitigate underdetermination problems, we set $\psi_D=0$ and tested two values of $C_D$, which are representative of values discussed in the literature~\cite{Sheinin:2011fv,Meng2014}.
For  $C_D=\SI{1}{nm}$ (red points), we find an island of optimal solutions characterized by $a_D>\SI{0.6}{nm}$ and $\lD$ close to $\SI{1}{nm}$.
For $C_D=\SI{28}{nm}$, we find $a_D\in [0.45-\SI{0.5}{nm}]$ and $\lD\in[2-\SI{5}{nm}]$. The dashed horizontal line in the figure indicates $\lD=\SI{2}{nm}$. \textbf{B}: Same analysis but for L-DNA. Here, we set $\psi_L=\SI{-0.5}{rad/bp}$ and test $C_L=\SI{10}{nm}$~\cite{Sheinin:2011fv,Oberstrass:2012jc}.
We find $a_L\in [0.5-\SI{0.6}{nm}]$ and $\ell_L\in [1-\SI{2}{nm}]$.
} 
\label{fig:aC}
\end{figure}

To estimate $\hat a$ and $\hlp$ for both D-DNA and L-DNA, we use the data of Sheinin et al.~\cite{Sheinin:2011fv} obtained for 2.2~kb long molecules, which consist of extension curves measured over a wide range of values of $\sigma$ (from -2.3 to +0.1) and $f$ (from 2.5 to \SI{36}{pN}). In accord with the findings of~\cite{Vlijm:2015yh}, we further consider that L-DNA occurs for $\sigma\lesssim-1$ and, hence, estimate D-DNA and L-DNA parameters using an optimization procedure of the extension curves for $\sigma$ lying in $[- 0.6,-0.1]$ and in $[-1.8,-1.3]$, respectively.
As discussed above, we lift the underdetermination problem by setting, on one hand, $\psi_D=0$ and $\psi_L=\SI{-0.5}{rad/bp}$ and we test, on the other hand, $C_D=\SI{1}{nm}$ as discussed in~\cite{Sheinin:2011fv}, $C_D=\SI{28}{nm}$ as proposed in~\cite{Meng2014} and $C_L=\SI{10}{nm}$ in accord with previous estimations~\cite{Sheinin:2011fv,Oberstrass:2012jc}.

Similar to the case of simulated data,
optimizations over forces from $2.5$ to $\SI{12}{pN}$ reveals the existence of small islands of optimal solutions, which we consider as our best estimations of parameters (Fig.~\ref{fig:aC}). For D-DNA (Fig.~\ref{fig:aC}A), $C_D=\SI{1}{nm}$ leads to $a_D>\SI{0.6}{nm}$ and $\lD$ close to $\SI{1}{nm}$, which are different, although being qualitatively similar, from previous estimations according to which $a_D=\SI{0.54}{nm}$ and $\lD\in[2-\SI{3}{nm}]$ (see~\cite{Sheinin:2011fv} and references therein). When $C_D=\SI{28}{nm}$, we find $a_D\in [0.45-\SI{0.5}{nm}]$ and $\lD\in[2-\SI{5}{nm}]$. Note, here, that we used the finest level of chain discretization ($n=1$) and we checked that our results remain identical for rougher coarse-graining up to $n=4$ (Fig.~S10).
The effect of $n$ is indeed expected to affect the exact value of the domain wall penalty $J$ (see above) and the entropic contribution to the free energy associated with the multiple possible positions of the denaturation bubbles (the term $\mathcal{J}(\la)$ in Eq.~\ref{eq:F}, see \mmet). However, as discussed using simulated data, the former is expected to play only a marginal role in the determination of $\hat a_D$ and $\hlD$, while in the case of a fixed, small number of denaturation bubbles as here (typically one), the latter scales as $\ln(n)$.

For L-DNA (Fig.~\ref{fig:aC}B), we find $a_L\in [0.5-\SI{0.6}{nm}]$, which is slightly larger than previous estimations~\cite{Sheinin:2011fv}, and $\lL\in [1-\SI{2}{nm}]$, which is slightly smaller. In all cases, the small value of $\lL$ corroborates that L-DNA is not a pure left-handed structural form but made of denatured DNA~\cite{Sheinin:2011fv,Oberstrass:2012jc}.

To summarize, our results question the compatibility of previously used $a_D=\SI{0.54}{nm}$ with reported torsional parameters. They also show the importance of having precise knowledge of a certain number of parameters to be able to fully exploit magnetic tweezers experiments in the high force regime. In absence of such prior information (as it is usually the case for real molecules), several strategies can be contemplated. For instance, working with an optimization procedure that is constrained by both torque curves and extension curves should narrow down the set of parameters leading to good fits. Even in this case, yet, having access to the fraction of the alternative form seems mandatory (see \mmet). Potential development along this line might come from the use of DNA binding proteins sensitive to alternative forms. Additional constraints on model parameters could also be imposed using cyclization data and conformation statistics from cryo-electron microscopy experiments.

\subsubsection{Extension of the model by including sequence effects}

In effect, our 2sRLC model bridges "kinkable" worm-like chain models and self-avoiding rod like chain models, which have been developed to respectively address the high flexibility of sharply bent DNA~\cite{Cong:2016te} and the superstructuring properties of supercoiled B-DNA molecules~\cite{Strick2003,vol2015book}. By doing so, it opens novel perspectives in the field of supercoiled DNA, not only to better rationalize magnetic tweezers experiments but also to predict behaviors of bacterial genomes {\it in vivo}. In this regard, including sequence effects, such as the tendency of AT-tracks and CpG-tracks to respectively favor DNA denaturation and left-handed forms~\cite{Mirkin:2008uj}, can be easily realized by using sequence dependent free energy formation costs of alternative forms (the $\gamma_{s_i}$ in Eq.~\ref{eq:Egeff}). As an example, we simulated the folding of a 1~kb long molecule where one single site ($i_{AT}$) had a much lower $\gamma^0_D$ than the other sites, thus mimicking the presence of an AT-track. Just as previously found in a more detailed, base resolution polymer model of DNA~\cite{Matek:2015gb}, we find that for a force $f=\SI{1}{pN}$  and a supercoiling density $\sigma=-0.06$, the site $i_{AT}$ is systematically denatured and serves as both nucleation and anchor points for a plectoneme (Fig.~S11).

\section{Conclusion}

More elaborate 2sRLC models can be contemplated by e.g.~including sequence effects at the level of DNA bending~\cite{Geggier:2010rf,Mitchell:2017ip}. In particular, by properly distinguishing intrinsic curvature from stiffness~\cite{Mitchell:2017ip}, one can expect to quantitatively address the structuring properties of bacterial sequences, including the possible pinning of pure B-DNA plectonemes~\cite{Kim:bd}. In all cases, our ability to develop truly predictive models of supercoiled DNA depends on our capacity to precisely infer mechanical parameters of alternative forms, which itself relies on the derivation of analytical expressions of thermodynamic quantities as a function of system parameters, at least for some range of forces and supercoiling densities~\cite{Sheinin:2011fv,Oberstrass:2012jc,Meng2014}. Here, we have derived simple analytical expressions that accurately describe the free energy landscape of the 2sRLC model in the approximation of zero writhe, providing a powerful alternative to phenomenological approaches. It would then be interesting to compare our analytical results to other exact analytical results obtained in models devoid of self-avoidance~\cite{Efremov:2016pr}. It would also be interesting to further expand the free energy landscape in power series of the writhe. This would allow including additional constraints coming from data at lower forces where alternative forms co-exist with superstructures~\cite{Meng2014}.

\r{This latter possibility appears to be crucial for unambiguously estimating parameters of alternative DNA forms. We have indeed shown that the proper estimation of a specific parameter may critically depend on the knowledge of other parameters or, equivalently, on their independent estimation  -- see e.g.~the impact of the torsional modulus on the estimation of both the distance between consecutive base pairs and the persistence length in Fig.~\ref{fig:aC}B. A systematic {\it parallel} analysis of extension curves together with torque curves (which is particularly adapted to highlight torsional properties) seems therefore necessary. In addition, sequences of interest, like GC or AAT repeats as implemented in ~\cite{Oberstrass:2012jc,Oberstrass:2013du}, could be systematically used to channel the formation of alternative forms, which would have two main advantages. On one hand, this would allow "controlling" the fraction of the alternative form ($\lambda$ in Eq.~\ref{eq:F}), thus lifting an important unknown of the problem. On the other hand, this would sharpen signals because of stronger cooperative transitions (see e.g.~\cite{Oberstrass:2013du}) and, hence, should constrain further model predictions -- just as for the co-existence of plectonemic conformations and strecthed conformations with alternative forms (Figs.~\ref{fig:simus} and S2), this may also require specific numerical methods to efficiently sample the different states at equilibrium. For more general sequences, DNA binding proteins sensitive to alternative forms, as used e.g.~in~\cite{Vlijm:2015yh}, might offer solutions to compute the fraction $\lambda$.}

\section*{Acknowledgements}
We thank Ruggero Cortini and Marc Joyeux for their early input in this project, Bahram Houchmandzadeh for useful suggestions, Daniel Jost for critical reading of the manuscript and Olivier Rivoire for helpful comments. We also thank Cees Dekker and John van Noort for providing us with their experimental data. I.J. is supported by an ATIP-Avenir grant (Centre National de la Recherche Scientifique).

\section*{Appendix A. Supplementary Material}

The supplementary material is a single pdf file consisting of two sections: I) the "Supplementary methods" section provides details related to the analytics and simulation of the 2sRLC model and II) the "Supplementary figures" section contains 11 figures.

\section*{Author Contributions}
I.J and T.L. performed research; T.L. implemented simulation tools; I.J. and T.L. analyzed data; I.J. wrote the article.



\section*{Supplementary material}

## II. SUPPLEMENTARY FIGURES

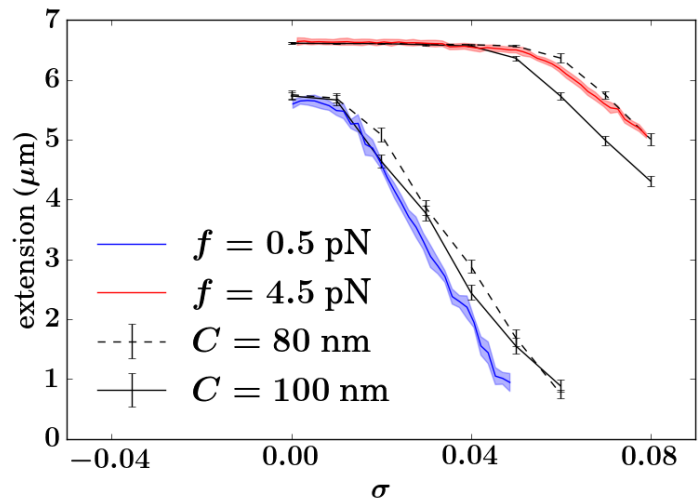

FIG. S1. Effect of $C$ on the simulated extension curves. Using $C = 100$ nm fits better the experimental data for $f = 0.5$ pN while $C = 80$ nm gives a better agreement for $f = 4.5$ pN.

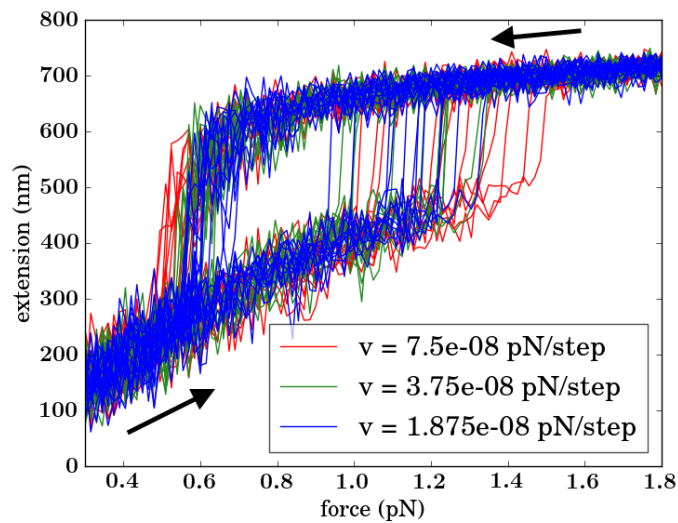

FIG. S2. Hysteresis cycles around the plectonemic-denatured transition. Simulations of a 2.4 kb molecule were performed ($\sigma = -0.05$) during which the force was continuously increased (as indicated by the lower left arrow) and decreased (upper right arrow arrow) at constant speed from 0.3 to 1.8 pN. For each of the three speeds reported here (from the fastest to the slowest: red, green and blue curves), 12 cycles were performed, revealing in each case a strong hysteresis pattern in the force-extension diagram.

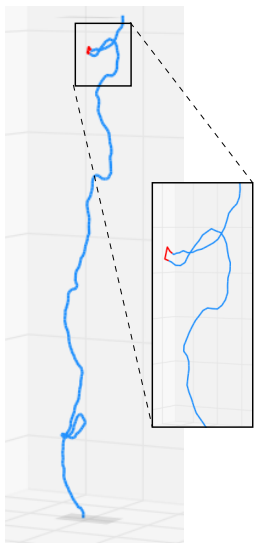

FIG. S3. Snapshot of a simulation close to the plectoneme-denaturation transition (in the figure S2). A denaturation bubble (red) appears at the apex of a plectoneme.

10

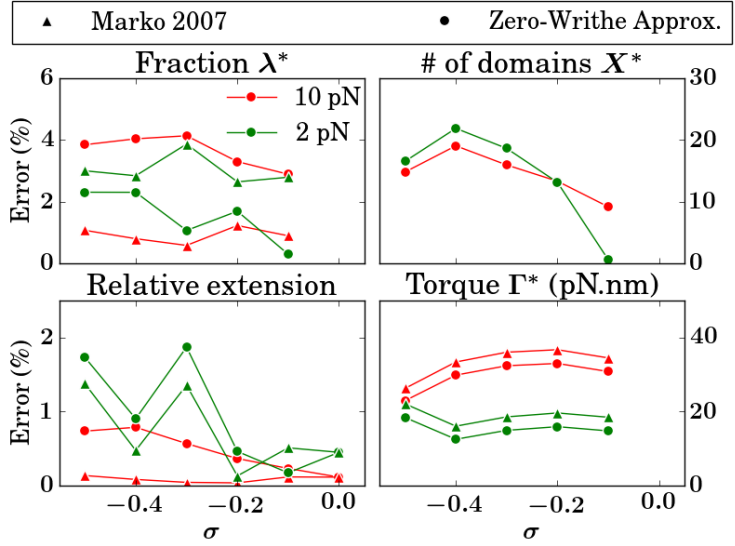

FIG. S4. Relative difference between the zero-writhe approximation and our simulations and between Marko's model and our simulations, corresponding to Fig. 3 in the main text.

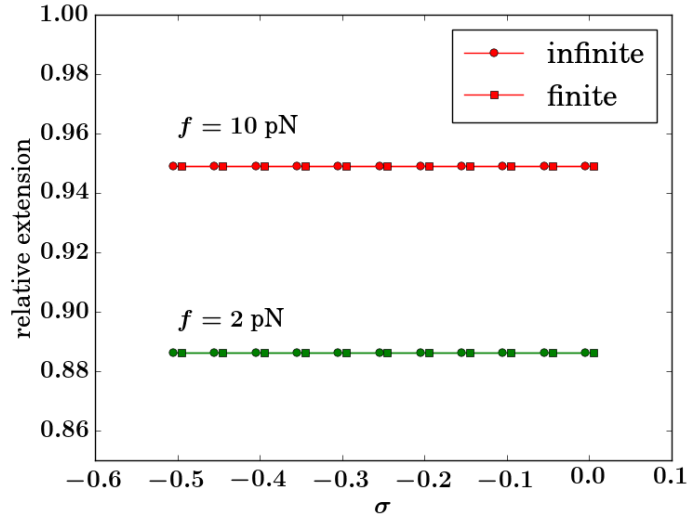

FIG. S5. FInite-size effects on the zero-writhe approximation: extension as a function of $\sigma$ for 2 different forces, including finite-size effects (squares) or not (circles). The points are actually almost exactly superimposed; they were slightly offset along the x-axis for readability.

11

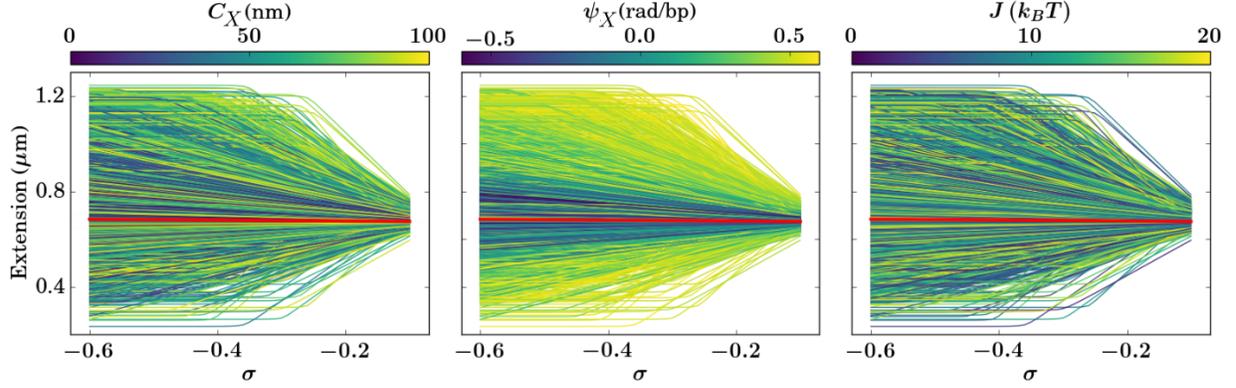

FIG. S6. Rainbow plots of the extension with respect to the varying parameters not shown in Fig.3 of the main text, namely, $C_X$, $\psi_X$ and $J$. $C_X$ and $J$ have no effect on the extension. For $\psi_X$, values away from $\psi_B = 0.6\,\mathrm{rad/bp}$ tend to yield better fits.

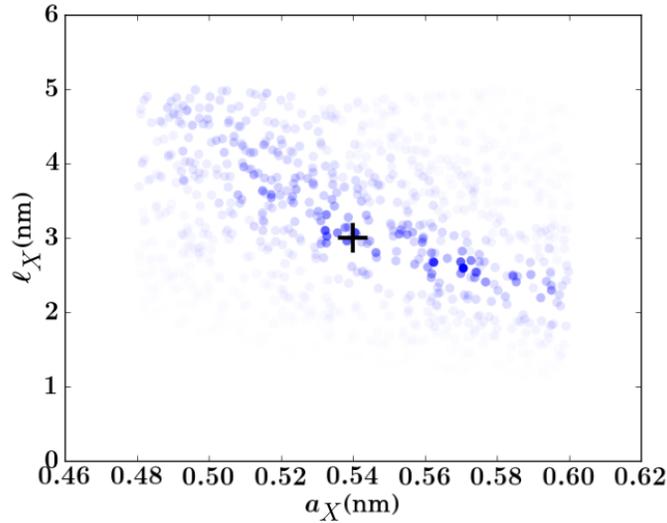

FIG. S7. In a situation where $\psi = \hat{\psi}_X$ and $C = \hat{C}_X$, combining RMSDs for multiple forces $(2, 2.5, 3, 3.5, 6, 8.5$ and $12\,\mathrm{pN})$, the best solutions resulting from the intersection of the crests shown in Fig. 2 of the main text are located around $(\hat{a}_X, \hat{\ell}_X)$ (black cross).

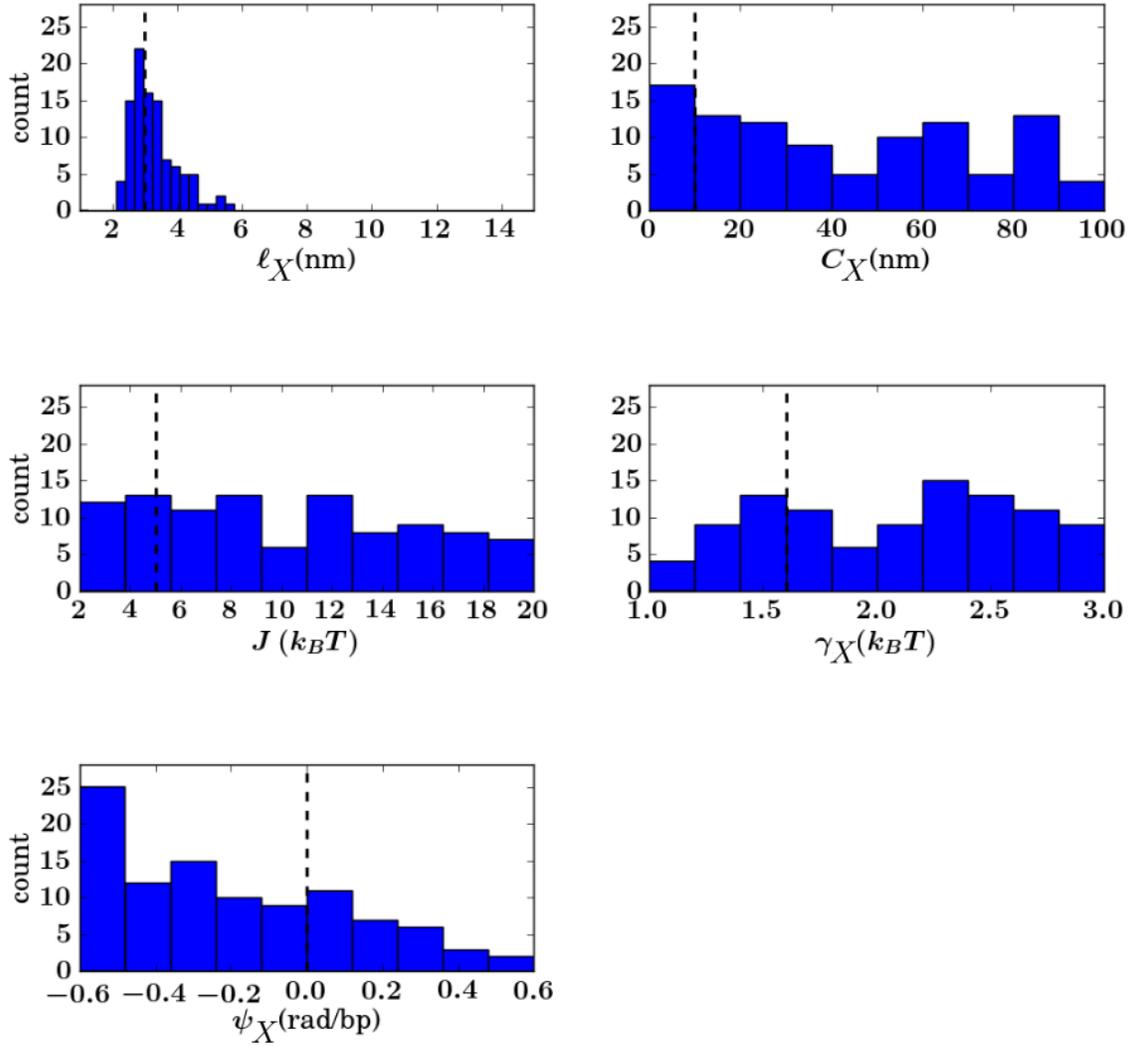

FIG. S8. Typical values of parameters for best fit solutions. Within the slice $a_X \in [0.52, 0.56\,\text{nm}]$ centered around $\hat{a}_X$ (see Fig. 4 in the main article), we selected the 100 best fits to the fake data and plotted the histogram of values for the other parameters (dashed lines represent $\hat{\ell}_X, \hat{C}_X, \hat{J}$ and $\hat{\gamma}_X$). Contrary to the other parameters that spread across all their allowed range, the values of $\ell_X$ are concentrated around $\hat{\ell}_X$ (top left).

13

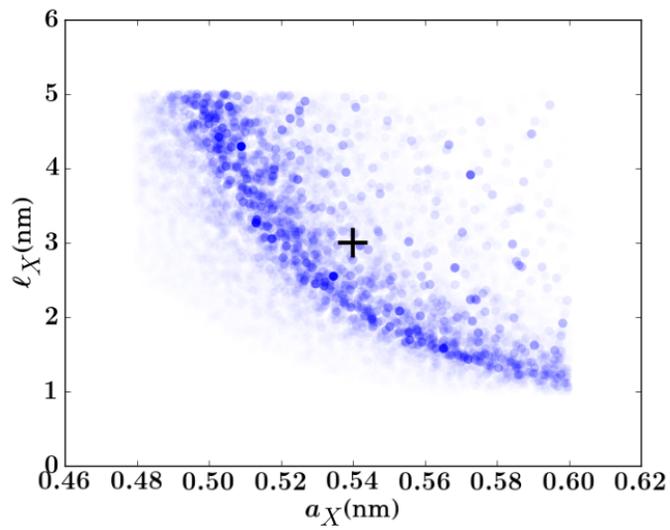

FIG. S9. Setting only $\psi_X = \hat{\psi}_X \ (= 0)$, at high force (12 pN) we find a crest of sub-optimal fits that go off the original parameters $(\hat{a}_X, \hat{\ell}_X)$ (black cross).

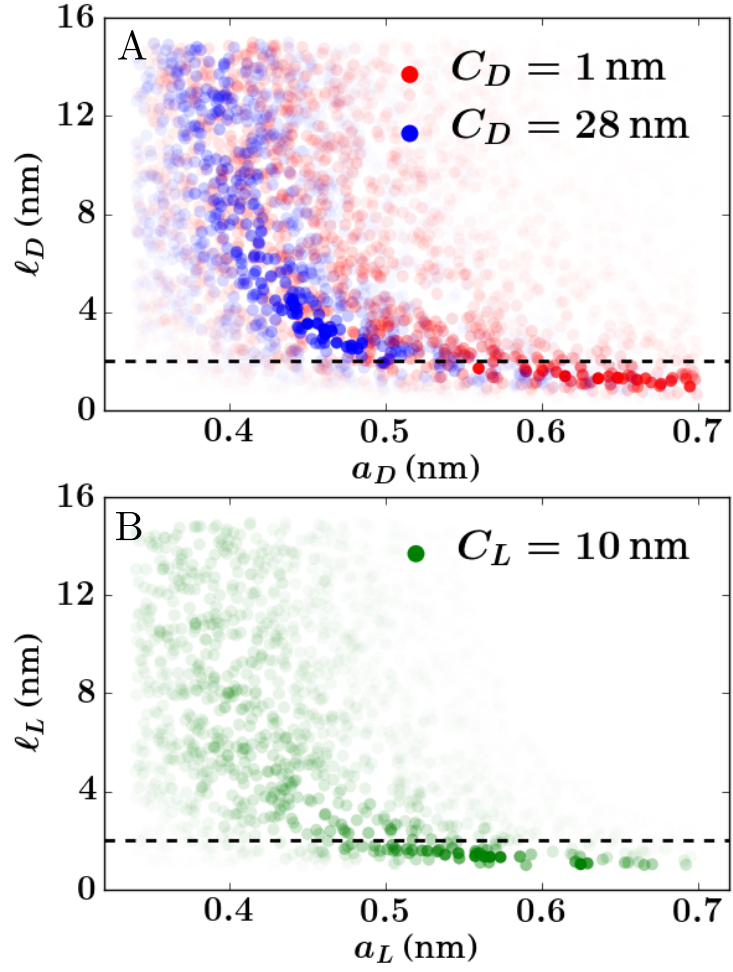

FIG. S10. Effect of the coarse-graining on the zero-writhe approximation. The results of Fig. 5 of the main text are unchanged when using a different level of discretization (here $n = 4$).

15

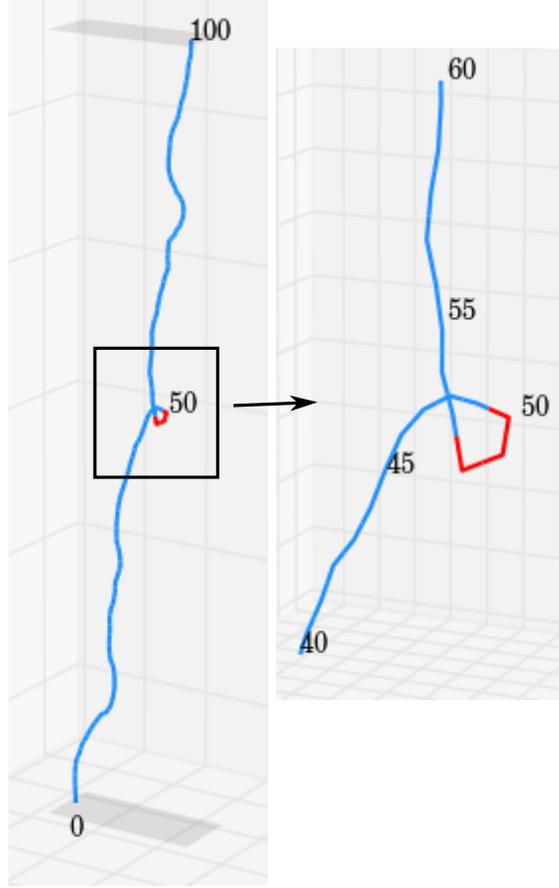

FIG. S11. Snapshot of a simulation of a 1 kb-long molecule (100 cylinders, $f = 1\,\text{pN}$, $\sigma = -0.06$) where the cylinder at the center (number 50) has no denaturation penalty ($\gamma_{D,i=50} = 0$). The denaturation bubble (red) appears and stays at the center during the whole simulation.



\end{document}